\documentclass{aa}  
\usepackage{graphicx}
\usepackage{txfonts}
\usepackage{natbib}
\bibpunct{(}{)}{;}{a}{}{,}
\begin{document}
   \title{The host galaxy of GRB\,031203: a new spectroscopic study}


   \author{R. Margutti\inst{1}
          \and
          G. Chincarini\inst{1,2}
          \and
          S. Covino\inst{2}
          \and
          G. Tagliaferri\inst{2}
          \and
          S. Campana\inst{2}
          \and 
          M. Della Valle\inst{3}
          \and
          A. V. Filippenko\inst{4}
          \and
          F. Fiore\inst{5}
          \and
          R. Foley\inst{4}
           \and
		D. Fugazza\inst{2}
          \and
          P. Giommi \inst{6}
          \and
          D. Malesani\inst{7}
          \and  
          A. Moretti\inst{2}
          \and
           L. Stella\inst{5}
                     }

   \offprints{R. Margutti}

   \institute{Universit\'a degli Studi di Milano-Bicocca, Dipartimento di Fisica, piazza della Scienza 3, I-20126 Milano, Italy.\\
              \email{raffaella.margutti@brera.inaf.it}
         \and INAF, Osservatorio Astronomico di Brera, via E. Bianchi 46, I-23807 Merate (Lc), Italy.
         \and INAF, Osservatorio Astrofisico di Arcetri, largo E. Fermi 5,I-50125 Firenze, Italy.
		 \and Department of Astronomy, University of California, Berkeley, CA 94720-3411.
         \and INAF, Osservatorio Astronomico di Roma, via Frascati 33, Monteporzio Catone (RM), I-00040, Italy.
         \and ASI, Science Data Center, ASDC, c/o ESRIN, Via G. Galilei, I-00044 Frascati, Italy.
		\and Dark Cosmology Centre, Niels Bohr Institute, University of Copenhagen, Juliane Maries Vej 30, DK-2100 K\o{}benhavn \O, Denmark.
		\\
}
\date{Received 23 April 2007; accepted 21 August 2007}

 
  \abstract
{}
  {The host galaxy of the long-duration gamma-ray burst (GRB) 031203 
(HG\,031203) offers a precious opportunity to study in detail the 
environment of a nearby GRB. The aim is to 
better characterize this galaxy and analyse the 
possible evolution with time of the spectroscopic quantities we derive.
}
  {We analyse HG\,031203 using a set of optical spectra acquired with 
the ESO-VLT and Keck telescope. We compare the metallicity, luminosity and star
formation properties of this galaxy and of the other supernova-long gamma-ray burst  hosts in the local universe ($z<0.2$)
against the KPNO International Spectroscopic Survey. }
      {
HG\,031203 is a metal poor, actively star forming
galaxy (star formation rate of $12.9\pm2.2\, \mathrm{M_{\sun} \,yr^{-1}}$) at $z=0.1054$.
From the emission-line analysis we derive an intrinsic reddening $E_{\mathrm{HG}}(B-V)\approx 0.4$. 
This parameter doesn't show a compelling evidence of evolution at a month time-scale. 
We find an interstellar medium
temperature of $\approx 12500\, \mathrm{K}$ and an electronic density of $N_{\mathrm{e}}=160\, \mathrm{cm^{-3}}$. 
After investigating for possible Wolf-Rayet emission features in our spectra,
we consider dubious the classification of HG\,031203 as a Wolf-Rayet galaxy.
Long gamma-ray burst (LGRB) and supernova hosts in the local universe ($z<0.2$) show, on average, specific
star formation rates higher than ordinary star forming galaxy at the same redshift.
On the other hand, we find that half of the hosts follows the metallicity-luminosity relation
found for star-burst galaxies;
HG031203 is a clear outlier, with its really low metallicity ($12+\log{(\mathrm{O/H})}=8.12\pm0.04$).
 }
  {}


   \maketitle


\section{Introduction}
While it is now widely believed that long gamma-ray bursts (LGRBs) arise from the core collapse of young massive stars
 (`\emph{collapsar\/}' model, 
\citealt{Woosley,FadyenWoosley}), the nature of their host galaxies (HGs) is still under discussion. 
The properties of about 50 HGs of LGRBs have been identified to date: 
 nevertheless, 
we cannot tell whether they form a new class of objects or just a sub-sample of an already known population of galaxies.

Among the physical properties of the circumburst environment, the 
metallicity seems to play a central role in the formation of these 
highly energetic events. The ``collapsar'' model favours progenitors
of low metallicity because of the reduced angular momentum loss and 
mass loss via strong winds at the surface of the star. Limited 
chemical evolution of GRB hosts is also the explanation suggested by 
\cite{Fruchter06} in order to account for the observed differences 
between LGRB HGs and HGs of core-collapse supernovae (CC~SNe):
while the rate of CC~SNe (per unit stellar luminosity) is almost 
equal in spiral and irregular galaxies (\citealt{Cappellaro};
\citealt{Leaman}), the overwhelming majority of LGRBs 
occur in faint, small irregulars. Moreover, CC~SN positions follow 
the blue light of their host galaxies within statistical uncertainties, 
while LGRBs are far more concentrated on the brightest HG regions 
\citep{Fruchter06}.  From the perspective of obtaining an accurate 
census of star formation in the universe, the preference of LGRBs for 
low-metallicity galaxies is a potential bias that limits the use of 
their hosts as cosmological tracers, at least at low redshift. However, spectroscopic 
measurements of the host metallicity in the redshift interval 
$0.4 < z < 1$ indicate that these objects follow the mass-metallicity 
relation recently found \citep{Savaglio05} for star-forming galaxies 
in the same redshift interval (\citealt{Savaglio06}).  
In this paper, we present a case study of the GRB host population: 
the host galaxy of the long-duration ($\sim 30\,\mathrm{s}$)
GRB\,031203 that triggered the IBIS instrument of INTEGRAL 
\citep{Gotz,Sazonov} on 2003 December 3.91769 
(UT dates are used throughout this paper). A small galaxy was 
found spatially coincident with the X-ray source \citep{Hsia} and 
was later identified as the GRB host. From optical spectroscopy, 
\cite{proc04} derived for HG\,031203 a redshift $z = 0.1055$.

While making GRB\,031203 one of the closest GRBs ever observed, 
the proximity of the burst also triggered particular interest in
attempting to detect an associated SN explosion. The host was 
therefore monitored independently by different groups, searching 
for an SN rebrightening \citep{Thomsen,Cobb,Gal}. The light curve 
of HG\,031203 indeed showed the typical photometric variability of 
an SN superimposed on a host galaxy. Ultimate confirmation came 
from the spectroscopic observations of \cite{Tagliaferri}; the event
was named SN\,2003lw. We refer to \cite{Malesani} for a complete 
discussion of the SN\,2003lw explosion and for details about 
HG\,031203 photometry.

The proximity of HG\,031203 allows us to perform a detailed analysis of 
its physical properties. From high-quality optical observations, 
we determine the host-galaxy metallicity, intrinsic reddening,
nebular parameters, SFR, and specific 
star-formation rate (SSFR).
Given the present set of data we can test the potential evolution of the host 
galaxy parameters on a month time scale.
We further characterize this object,
directly contrasting its physical properties with the KPNO 
International Spectroscopic Survey (KISS), an emission-line 
survey of galaxies in the local universe ($z \leq 0.1$) that
utilized low-dispersion objective-prism spectra \citep{Salzer}.

\begin{figure*}[t]
\centering
\includegraphics[ scale=1 ]{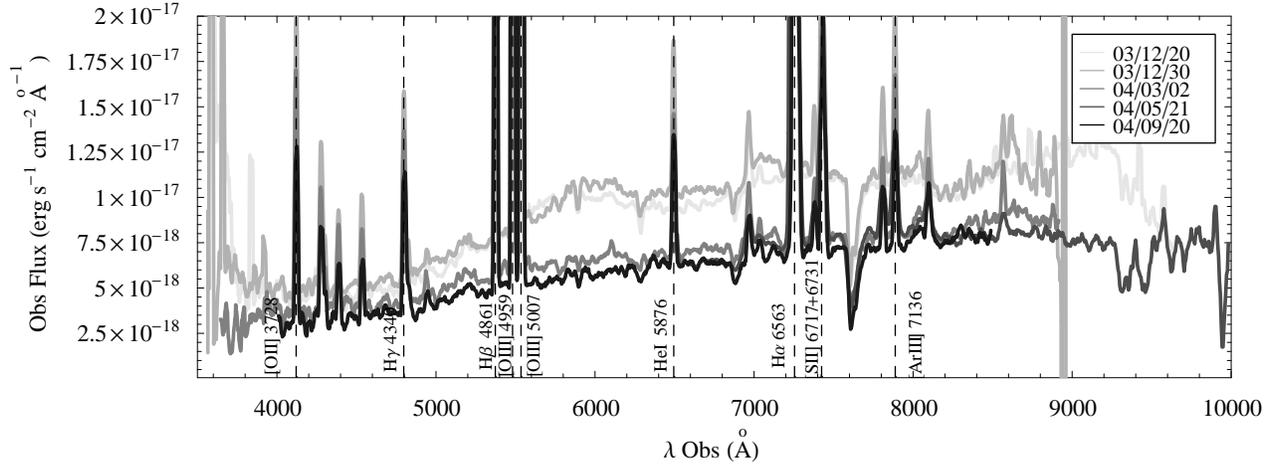}
\caption{VLT spectra of the host galaxy of GRB\,031203. The different continuum emission is due to the contribution
of SN2003lw. Note the strong $\mathrm{H\alpha}$ emission, corresponding to a SFR $\approx 13\,\mathrm{M_{\sun}\, yr^{-1}}$.  
No extinction correction has been applied.} 
\label{Figure1}
\end{figure*}

In order to better understand the nature of galaxies that are able to
produce one of the most energetic events in the universe, we consider
the largest sample of local LGRB hosts with associated SN
explosions. This sample includes HG\,980425 \citep[$z =
0.0085$;][]{Galama}, HG\,030329 \citep[$z =
0.1685$;][]{Sollerman05,Gorosabel,Thoene}, HG\,031203 ($z = 0.10536$,
this work), and HG\,060218 \citep[$z =
0.0335$;][]{Modjaz,Wiersema}. We focus on the metallicity and 
star-formation properties of these ga\-la\-xies, with the aim of shedding 
light on the possible peculiarity of these objects.

This paper is organised as follows. Section \ref{Sec:Obs} reviews 
the observations and shortly describes the performed data reduction. 
We assess the intrinsic and Galactic extinction toward GRB\,031203 
in Sect. \ref{Sec:Ext}, where the host is characterized in terms of 
emission-line fluxes. We perform an emission-line analysis to 
determine the host relative chemical abundances and star-formation 
properties in Sect. \ref{Sec:Met} and \ref{Sec:SFR}, respectively. 
Section \ref{Sec:Cont} analyses the continuum emission of HG\,031203, 
while in Sect. \ref{Sec:WR} we carefully investigate the possible 
presence of Wolf-Rayet (WR) features in the spectra of HG\,031203. 
Our findings are discussed and summarized in Sect. \ref{Sec:Disc}.
Finally, conclusions are drawn in Sect. \ref{Sec:Conc}.

Throughout this paper standard $\Lambda$ cosmology is used: $H_{0}=72\,\mathrm{km\, s^{-1}\,Mpc^{-1}}$,\, $\Omega _{\Lambda}=0.7$,
 \,$\Omega _{m}=0.3$.

\section{Observations and data analysis}
\label{Sec:Obs}

\emph{Photometry.---}We acquired $U$, $B$, and $V$ photometry of 
HG\,031203 with VLT+FORS2 during the night of 2004 March 26th 
(that is, about 4 months after the GRB). We expect negligible 
contribution from SN\,2003lw at this epoch (see, e.g.,
\citealt{Mazzali}). Observing conditions were
photometric. Data reduction was performed following standard procedures.
Flux ca\-li\-bra\-tion was achieved by observing the \cite{Landolt} standard field
Rubin\,152. The photometric calibration was carried out including
appropriate color terms, resulting in a systematic uncertainty of 
$\sim 0.05$~mag. The host galaxy was clearly detected in all filters, 
and its magnitudes are reported in Table \ref{Tab:Mag}. For
the extinction correction we used a Galactic color excess $E_{\rm{MW}}(B-V)=\,0.72$ and 
an intrinsic reddening
$E_{\rm{HG}}(B-V)=\,0.38$, value measured from the spectrum
taken on March 2nd, about 20 days before the photometric
observations (see Sect. \ref{Sec:Ext}). $K$-corrections were derived
interpolating the extinction corrected fluxes at the required wavelenghts. 
In particular we assessed the $I$-band $K$-correction 
considering the spectral slope of the spectrum taken on March 2nd.

From the badly constrained Galactic color excess (see Sect. \ref{Sec:Ext})
we derive additional systematic errors of 
$0.32$, $0.22$, $0.20$, $0.15$ and $0.10$ mag for 
the $U$ $B$ $V$ $R$ and $I$-band extinction corrected absolute
magnitudes respectively.

\begin{table}
\begin{minipage}[t]{\columnwidth}
\caption{Observing log (photometry). Middle exposures dates are provided.}
\label{Tab:ObsFot}
\centering
\begin{center}
\begin{tabular}{lllll}
\hline
Date & Band & Exp & Seeing  & Instrument  \\
(UT) &  &(s)& ($''$)& \\
\hline
\hline
2004 Mar 27.1044&$V$ &2$\times$120  & 1.1& VLT+FORS2\\
2004 Mar 27.0958&$B$ &2$\times$300 & 0.9&VLT+FORS2\\
2004 Mar 27.0835&$U$ &3$\times$300 & 1.1&VLT+FORS2\\
\hline
\end{tabular}
\end{center}
\end{minipage}
\end{table} 

\begin{table}
\begin{minipage}[t]{\columnwidth}
\caption{Observed magnitudes and extinction and $K$-corrected absolute magnitudes of HG\,031203.
The errors reported account for uncertainties coming from the flux calibration procedure
and the intrinsic extinction correction.}
\label{Tab:Mag}
\centering
\begin{center}
\begin{tabular}{lllll}
\hline
  Band & \multicolumn{2}{c}{Obs. mag.} & \multicolumn{2}{c}{ Abs. Mag.} \\
   & \multicolumn{2}{c}{(mag)}&\multicolumn{2}{c}{(mag)}\\
\hline
\hline
$U$& 22.31 & $\pm 0.18$ & $-21.38$ &$\pm$ 0.47\\
$B$& 22.32&$\pm 0.05$& $-21.00 $&$\pm$ 0.39\\
$V$& 20.53&$\pm 0.05$&$-21.02$ &$\pm$ 0.31\\
$R^\mathrm{a}$& 20.44&$\pm 0.02$&$-20.54$ &$\pm$ 0.24\\
$I^\mathrm{a}$& 19.40&$\pm 0.04$&$-19.90$ &$\pm$ 0.16\\
\hline
\end{tabular}
\end{center}
\end{minipage}
\begin{list}{}{}
\item[$^{\mathrm{a}}$] From \cite{Mazzali}.
\end{list}
 \end{table} 
 
 \emph{Spectroscopy.---}We acquired optical spectra of the host galaxy
of GRB\,031203 with the European Southern Observatory Very Large
Telescope (ESO-VLT, Cerro Paranal, Chile) and the Keck-I telescope
(Hawaii, USA).
 
 We first observed HG\,031203 on 2003 December 20.29 using the FORS2
(focal reducer/low-resolution spectrograph) instrument on VLT-UT4
(Yepun), characterized by a spatial resolution of
$0.126''\,\mathrm{pix^{-1}}$. With the 300V grism, the inverse
dispersion was $112\,\mathrm{\AA\, mm^{-1}} $ ($1.68\,\mathrm{ \AA\,
pix^{-1}}$), while we covered the wavelength range 3300--8700\,
\AA. Owing to the excellent seeing on that night (0.5$''$), the two
brightest emission lines ($\mathrm{H\alpha}$ and [O~III]
$\lambda$5007) are slightly saturated.

In order to extend the spectral coverage up to 11,000\, \AA, we also
acquired spectra using the 150I grism along with the order-blocking
filter OG590. The inverse dispersion was $3.45\,\mathrm{ \AA\,
pix^{-1}}$ ($230\,\mathrm{ \AA\, mm^{-1}}$). After calibration, we
compared the fluxes in the wavelength region common to the two grisms
(300V and 150I) and found good agreement. All our observations were
split into shorter exposures (see Table ~\ref{Tab:Obs}), and
Fig. \ref{Figure1} shows for each night the co-added spectrum.

Our second run was performed on 2003 December 30.22 u\-sing the FORS1 on
VLT-UT1 (Antu). This instrument is characterized by a lower spatial
resolution ($0.2''\,\mathrm{pix^{-1}}$). The inverse dispersion is
worse, too: $2.64\,\mathrm{ \AA \, pix^{-1}}$ and $5.5\,\mathrm{ \AA\,
pix^{-1}}$ for the 300V and 150I grisms, respectively. During that
night the seeing was 0.6$''$; no emission lines were saturated.

We acquired a first reference host-galaxy spectrum on 2004 March 02.09; at
that time we expected SN\,2003lw to have largely faded away, leaving a
negligible contribution to the total flux. This spectrum was acquired
with FORS1 + grism 300V, and it allowed us to evaluate the supernova
contribution in earlier spectra.

On 2004 March 16th, HG\,031203 was observed with the $10\,\mathrm{ m}$
Keck-I telescope. Spectra were acquired using the low re\-so\-lu\-tion
imaging spectrometer (LRIS; \citealt{Oke}) and the D560 dichroic.
Unfortunately, the night was characterized by poor seeing conditions
(about $2.3''$).

Additional reference 150I spectra were obtained on 2004 May 21 with
the FORS1 instrument on VLT-UT1. This observation confirmed that the
SN contribution was negligible in March ($\sim 0.03\,\mathrm{ mag}$).

Our last HG\,031203 spectra date back to more than 9 months after the
trigger of the burst: on 2004 September 20, we observed the host
galaxy with FORS2, grism 300V on VLT-UT4. The seeing was very good
($0.6''$), though no emission lines show saturation.

Each of the VLT spectra of HG\,031203 was obtained using a long slit
$1''$ wide, but the slit used for the standard stars LTT3864 and
LTT1788 was $5''$ wide.  Keck spectra were instead acquired using a
$1.5''$ slit.  All nights of observation were photometric.
The complete observing log is given in Table ~\ref{Tab:Obs}. \\

\begin{table}
\begin{minipage}[t]{\columnwidth}
\caption{Observing log (spectroscopy). Middle exposures dates are provided.}
\label{Tab:Obs}
\centering
\begin{center}
\begin{tabular}{lllll}
\hline
Date & Grism & Exp & Seeing  & Instrument  \\
(UT) &  &(s)& ($''$)&\\
\hline
\hline
2003 Dec 20.29 & 150I & 900 & 0.5 & VLT+FORS2 \\
2003 Dec 20.30 & 150I & 900 & 0.5 & VLT+FORS2 \\
2003 Dec 20.33 & 300V  & 2700 & 0.5 &  VLT+FORS2 \\
2003 Dec 20.36 & 300V  & 2700 & 0.5 &  VLT+FORS2 \\
2003 Dec 30.22 & 150I & 900 & 0.5 & VLT+FORS1 \\
2003 Dec 30.23 & 150I & 900 & 0.5 & VLT+FORS1 \\
2003 Dec 30.28 & 300V  & 2700 & 0.5 &  VLT+FORS1 \\
2003 Dec 30.32 & 300V  & 2700 & 0.5 &  VLT+FORS1 \\
2004 Mar 02.09 & 300V  & 1800 & 06 & VLT+FORS1 \\
2004 Mar 02.11 & 300V  & 1800 & 06 & VLT+FORS1 \\
2004 Mar 02.13 & 300V  & 1800 & 06 & VLT+FORS1 \\
2004 Mar 02.16 & 300V  & 1800 & 06 & VLT+FORS1 \\
2004 Mar 16.33 & 400 I\footnote{Grating.}&	1200 &	2.3	& Keck+LRIS+D560 \\
2004 Mar 16.35 & 600 B$^{a}$&	1200 &	2.3	& Keck+LRIS+D560 \\
2004 May 21.98 & 150I & 900 & 0.6 & VLT+FORS1 \\
2004 May 21.99 & 150I & 900 & 0.6 & VLT+FORS1 \\
2004 May 22.01 & 150I & 900 & 0.6 & VLT+FORS1 \\
2004 Sept 20.35 & 300V & 1800 &	0.6	& VLT+FORS2 \\
2004 Sept 20.37 & 300V & 1800 &	0.6	& VLT+FORS2 \\
\hline
\end{tabular}
\end{center}
\end{minipage}
\end{table}
\begin{table*}
\begin{minipage}[t]{\columnwidth}
\caption{Emission line summary (VLT observations).}
\label{Tab:Obs1}
\centering
\renewcommand{\footnoterule}{}  
\begin{tabular}{lr|rrrrr|rrrrrr}
\hline 
& &\multicolumn{5}{|c|}{2003 December 20th}&\multicolumn{5}{|c}{2003 December 30th}\\
\cline{3-12}
$\lambda_{\mathrm{rest}}$ &Ion    &$EW_{\mathrm{rest}}$&\multicolumn{2}{c}{Obs. Flux} 	&\multicolumn{2}{c|}{Ext. Corr. Flux} & $EW_{\mathrm{rest}}$&\multicolumn{2}{c}{Obs. Flux} 	&\multicolumn{2}{c}{Ext. Corr. Flux}\\
(\AA) 		     & 	     &(\AA)  	& \multicolumn{2}{c}{($10^{-17}\,\mathrm{cgs}$)}  &\multicolumn{2}{c|}{($10^{-17}\,\mathrm{cgs}$)}& (\AA)  	& \multicolumn{2}{c}{($10^{-17}\,\mathrm{cgs}$)}  &\multicolumn{2}{c}{($10^{-17}\,\mathrm{cgs}$)} \\
\hline\hline
3728.8	&	[OII]	&	52.04	&	26.65	&$\pm$	0.75	&	2718.66	&$\pm$	543.92	&	61.74	&	34.23	&$\pm$	2.49	&	2843.19	&$\pm$	571.69	\\
3868.7	&	[NeIII]	&	32.23	&	16.06	&$\pm$	0.45	&	1407.65	&$\pm$	275.24	&	32.01	&	17.69	&$\pm$	1.29	&	1269.62	&$\pm$	250.16	\\
3888.9	&	HeI	&	12.31	&	6.11	&$\pm$	0.17	&	523.63	&$\pm$	102.01	&	10.99	&	6.07	&$\pm$	0.44	&	426.31	&$\pm$	83.73	\\
3967.5	&	[NeIII]	&	18.01	&	8.80	&$\pm$	0.25	&	688.15	&$\pm$	132.05	&	16.92	&	9.32	&$\pm$	0.68	&	599.77	&$\pm$	116.24	\\
4101.7	&	HI	&	14.44	&	8.19	&$\pm$	0.23	&	547.38	&$\pm$	102.03	&	18.83	&	10.76	&$\pm$	0.78	&	595.35	&$\pm$	112.49	\\
4340.4	&	HI  	&	35.02	&	19.58	&$\pm$	0.55	&	989.67	&$\pm$	173.78	&	36.77	&	22.48	&$\pm$	1.64	&	950.82	&$\pm$	170.61	\\
4363.2	&	[OIII]	&	6.05	&	3.45	&$\pm$	0.10	&	170.01	&$\pm$	29.67	&	6.74	&	4.18	&$\pm$	0.30	&	172.51	&$\pm$	30.79	\\
4471.5	&	HeI	&	3.43	&	2.20	&$\pm$	0.06	&	96.04	&$\pm$	16.28	&	2.74	&	1.90	&$\pm$	0.14	&	69.87	&$\pm$	12.16	\\
4861.3	&	HI	&	83.15	&	71.70	&$\pm$	2.03	&	2142.39	&$\pm$	326.70	&	95.63	&	83.45	&$\pm$	6.08	&	2135.91	&$\pm$	340.46	\\
4959.0	&	[OIII]	&	184.51	&	171.58	&$\pm$	4.85	&	4724.45	&$\pm$	702.59	&	210.11	&	192.84	&$\pm$	14.04	&	4566.96	&$\pm$	713.23	\\
5006.9	&	[OIII]	&	561.02	&	526.73	&$\pm$	14.90	&	13958.94	&$\pm$	2051.15	&	634.64	&	596.61	&$\pm$	43.43	&	13623.87	&$\pm$	2107.08	\\
5875.6	&	HeI  	&	15.44	&	15.98	&$\pm$	0.45	&	244.34	&$\pm$	29.99	&	16.58	&	18.52	&$\pm$	1.35	&	250.52	&$\pm$	33.68	\\
6300.3	&	[OI]	&	4.09	&	4.72	&$\pm$	0.13	&	57.58	&$\pm$	6.57	&	4.48	&	5.57	&$\pm$	0.41	&	60.72	&$\pm$	7.74	\\
6312.1	&	[SIII]  	&	2.77	&	3.21	&$\pm$	0.09	&	38.91	&$\pm$	4.43	&	2.28	&	2.88	&$\pm$	0.21	&	31.21	&$\pm$	3.97	\\
6363.8	&	[OI]	&	1.87	&	2.21	&$\pm$	0.06	&	26.14	&$\pm$	2.95	&	...	&	...	&	...	&	...	&	...	\\
6548.1	&	[NII]	&	9.62	&	11.88	&$\pm$	0.43	&	127.43	&$\pm$	14.21	&	7.38	&	9.51	&$\pm$	0.82	&	91.67	&$\pm$	12.08	\\
6562.9	&	HI	&	459.85	&	569.15	&$\pm$	20.52	&	6060.99	&$\pm$	674.36	&	484.55	&	621.59	&$\pm$	53.47	&	5944.68	&$\pm$	782.40	\\
6583.4	&	[NII]  	&	25.18	&	31.24	&$\pm$	1.13	&	329.19	&$\pm$	36.50	&	26.45	&	33.79	&$\pm$	2.91	&	319.86	&$\pm$	42.01	\\
6678.2	&	 HeI 	&	6.58	&	6.47	&$\pm$	0.18	&	64.90	&$\pm$	6.93	&	4.16	&	7.94	&$\pm$	0.58	&	71.70	&$\pm$	8.72	\\
6716.5	&	[SII]  	&	15.40	&	19.00	&$\pm$	0.54	&	186.82	&$\pm$	19.81	&	18.58	&	23.27	&$\pm$	1.69	&	206.04	&$\pm$	24.94	\\
6730.7	&	[SII]	&	12.43	&	15.28	&$\pm$	0.43	&	149.10	&$\pm$	15.77	&	14.19	&	17.74	&$\pm$	1.29	&	155.99	&$\pm$	18.85	\\
7065.3	&	HeI	&	6.81	&	8.13	&$\pm$	0.23	&	66.76	&$\pm$	6.63	&	8.42	&	10.51	&$\pm$	0.77	&	78.32	&$\pm$	9.07	\\
7135.8	&	[AIII]	&	15.36	&	18.50	&$\pm$	0.52	&	146.66	&$\pm$	14.37	&	15.49	&	19.49	&$\pm$	1.42	&	140.37	&$\pm$	16.12	\\
7281.3	&	HeI	&	1.34	&	1.56	&$\pm$	0.04	&	11.53	&$\pm$	1.10	&	...	&	...	&	...	&	...	&	...	\\
7319.9	&	[OII]	&	4.09	&	4.64	&$\pm$	0.13	&	33.56	&$\pm$	3.17	&	5.82	&	6.97	&$\pm$	0.51	&	45.95	&$\pm$	5.15	\\
7330.2	&	[OII]	&	3.93	&	4.43	&$\pm$	0.13	&	31.86	&$\pm$	3.00	&	2.89	&	3.44	&$\pm$	0.25	&	22.54	&$\pm$	2.52	\\
7751.0	&	[AIII] 	&	4.38	&	5.51	&$\pm$	0.16	&	32.64	&$\pm$	2.82	&	3.52	&	4.65	&$\pm$	0.34	&	25.33	&$\pm$	2.69	\\
\hline
& &\multicolumn{5}{|c|}{2004 March 2nd}&\multicolumn{5}{|c}{2004 September 20th}\\
\cline{3-12}
$\lambda_{\mathrm{rest}}$ &Ion    &$EW_{\mathrm{rest}}$&\multicolumn{2}{c}{Obs. Flux} 	&\multicolumn{2}{c|}{Ext. Corr. Flux} & $EW_{\mathrm{rest}}$&\multicolumn{2}{c}{Obs. Flux} 	&\multicolumn{2}{c}{Ext. Corr. Flux}\\
(\AA) 		     & 	     &(\AA)  	& \multicolumn{2}{c}{($10^{-17}\,\mathrm{cgs}$)}  &\multicolumn{2}{c|}{($10^{-17}\,\mathrm{cgs}$)}& (\AA)  	& \multicolumn{2}{c}{($10^{-17}\,\mathrm{cgs}$)}  &\multicolumn{2}{c}{($10^{-17}\,\mathrm{cgs}$)} \\
\hline \hline
3728.8	&	[OII]	&	70.77	&	28.41	&$\pm$	2.85	&	2325.44	&$\pm$	885.80	&	72.33	&	23.85	&$\pm$	1.22	&	2157.33	&$\pm$	373.52	\\
3868.7	&	[NeIII]	&	33.49	&	14.20	&$\pm$	1.43	&	1004.50	&$\pm$	374.40	&	40.04	&	13.76	&$\pm$	0.70	&	1072.90	&$\pm$	181.84	\\
3888.9	&	HeI	&	11.71	&	5.00	&$\pm$	0.50	&	346.29	&$\pm$	128.63	&	14.08	&	4.87	&$\pm$	0.25	&	371.16	&$\pm$	62.69	\\
3967.5	&	[NeIII]	&	21.83	&	9.60	&$\pm$	0.97	&	608.75	&$\pm$	222.90	&	22.36	&	7.90	&$\pm$	0.40	&	551.02	&$\pm$	91.78	\\
4101.7	&	HI	&	21.38	&	9.06	&$\pm$	0.91	&	494.01	&$\pm$	175.99	&	22.00	&	8.43	&$\pm$	0.43	&	504.45	&$\pm$	81.79	\\
4340.4	&	HI	&	42.53	&	20.09	&$\pm$	2.02	&	838.25	&$\pm$	282.35	&	47.28	&	19.33	&$\pm$	0.99	&	879.95	&$\pm$	135.07	\\
4363.2	&	[OIII]	&	7.40	&	3.53	&$\pm$	0.35	&	143.71	&$\pm$	48.13	&	8.05	&	3.31	&$\pm$	0.17	&	146.98	&$\pm$	22.44	\\
4471.5	&	HeI	&	...	&	...	&	...	&	...	&	...	&	...	&	...	&	...	&	...	&	...	\\
4861.3	&	HI	&	117.02	&	76.39	&$\pm$	7.68	&	1932.24	&$\pm$	571.15	&	129.95	&	71.46	&$\pm$	3.64	&	1949.33	&$\pm$	263.49	\\
4959.0	&	[OIII]	&	260.63	&	179.06	&$\pm$	18.00	&	4192.25	&$\pm$	1211.10	&	278.66	&	165.38	&$\pm$	8.43	&	4167.29	&$\pm$	550.92	\\
5006.9	&	[OIII]	&	784.34	&	551.31	&$\pm$	55.41	&	12448.40	&$\pm$	3557.12	&	870.54	&	513.48	&$\pm$	26.18	&	12467.11	&$\pm$	1630.82	\\
5875.6	&	HeI	&	22.18	&	16.67	&$\pm$	1.68	&	223.34	&$\pm$	54.33	&	26.79	&	18.97	&$\pm$	0.97	&	270.05	&$\pm$	30.26	\\
6300.3	&	[OI]	&	6.27	&	5.26	&$\pm$	0.53	&	56.86	&$\pm$	12.98	&	6.46	&	4.87	&$\pm$	0.25	&	55.65	&$\pm$	5.87	\\
6312.1	&	[SIII]	&	3.31	&	2.78	&$\pm$	0.28	&	29.89	&$\pm$	6.81	&	1.55	&	1.18	&$\pm$	0.06	&	13.43	&$\pm$	1.41	\\
6363.8	&	[OI]	&	2.53	&	2.18	&$\pm$	0.22	&	22.85	&$\pm$	5.17	&	...	&	...	&	...	&	...	&	...	\\
6548.1	&	[NII]	&	9.77	&	8.51	&$\pm$	1.04	&	81.40	&$\pm$	18.78	&	12.77	&	9.73	&$\pm$	0.52	&	98.07	&$\pm$	10.13	\\
6562.9	&	HI	&	655.58	&	569.14	&$\pm$	69.47	&	5396.04	&$\pm$	1242.59	&	757.17	&	577.85	&$\pm$	31.12	&	5776.77	&$\pm$	595.67	\\
6583.4	&	[NII]	&	33.30	&	28.75	&$\pm$	3.51	&	269.81	&$\pm$	61.96	&	37.45	&	28.62	&$\pm$	1.54	&	283.15	&$\pm$	29.11	\\
6678.2	&	[HeI]	&	8.58	&	7.19	&$\pm$	0.72	&	64.36	&$\pm$	13.89	&	8.81	&	6.77	&$\pm$	0.35	&	63.84	&$\pm$	6.38	\\
6716.5	&	[SII]	&	25.39	&	21.03	&$\pm$	2.11	&	184.68	&$\pm$	39.63	&	27.47	&	21.16	&$\pm$	1.08	&	195.65	&$\pm$	19.46	\\
6730.7	&	[SII]	&	20.73	&	17.11	&$\pm$	1.72	&	149.16	&$\pm$	31.94	&	22.23	&	17.14	&$\pm$	0.87	&	157.31	&$\pm$	15.61	\\
7065.3	&	HeI	&	10.52	&	9.16	&$\pm$	0.92	&	67.73	&$\pm$	13.77	&	10.82	&	8.57	&$\pm$	0.44	&	66.48	&$\pm$	6.28	\\
7135.8	&	[AIII]	&	21.71	&	18.95	&$\pm$	1.90	&	135.43	&$\pm$	27.22	&	20.70	&	16.60	&$\pm$	0.85	&	124.37	&$\pm$	11.63	\\
7281.3	&	HeI	&	...	&	...	&	...	&	...	&	...	&	...	&	...	&	...	&	...	&	...	\\
7319.9	&	[OII]	&	6.06	&	5.46	&$\pm$	0.55	&	35.75	&$\pm$	6.97	&	6.05	&	5.08	&$\pm$	0.26	&	34.77	&$\pm$	3.16	\\
7330.2	&	[OII]	&	5.03	&	4.45	&$\pm$	0.45	&	29.01	&$\pm$	5.65	&	4.58	&	3.83	&$\pm$	0.20	&	26.10	&$\pm$	2.37	\\
7751.0	&	[AIII]	&	5.24	&	5.33	&$\pm$	0.54	&	28.86	&$\pm$	5.24	&	...	&	...	&	...	&	...	&	..	\\
\hline
\end{tabular}
\end{minipage}
\end{table*}

\begin{table*}
\begin{center}
\begin{minipage}[t]{\columnwidth}
\caption{Emission line summary (Keck observations).}
\label{Tab:Obs2}
\centering
\renewcommand{\footnoterule}{}  
\begin{tabular}{lr|rrrrr}
\hline
& &\multicolumn{5}{c}{2004 March 16th} \\
\cline{3-7}
$\lambda_{\mathrm{rest}}$ &Ion    &$EW_{\mathrm{rest}}$&\multicolumn{2}{c}{Obs. Flux} 	&\multicolumn{2}{c}{Ext. Corr. Flux}\\
(\AA) 		     & 	     &(\AA)  	& \multicolumn{2}{c}{($10^{-17}\,\mathrm{cgs}$)}  &\multicolumn{2}{c}{($10^{-17}\,\mathrm{cgs}$)}\\
\hline
\hline
3728.8	&	[OII]	&	46.52	&	27.78	&$\pm$	7.84	&	2307.33	&$\pm$	1370.92	\\
3868.7	&	[NeIII]	&	26.97	&	14.10	&$\pm$	1.44	&	1012.15	&$\pm$	527.31	\\
3888.9	&	HeI	&	10.00	&	5.18	&$\pm$	0.52	&	364.01	&$\pm$	188.85	\\
3967.5	&	[NeIII]	&	16.27	&	8.59	&$\pm$	0.83	&	552.53	&$\pm$	281.95	\\
4101.7	&	HI	&	19.05	&	10.20	&$\pm$	0.96	&	564.08	&$\pm$	279.47	\\
4340.4	&	HI  	&	33.87	&	16.92	&$\pm$	1.77	&	715.57	&$\pm$	335.85	\\
4363.2	&	[OIII]	&	7.63	&	3.78	&$\pm$	0.40	&	155.76	&$\pm$	72.73	\\
4471.5	&	HeI	&	...	&	...	&	...	&	...	&	...	\\
4861.3	&	HI	&	94.18	&	76.15	&$\pm$	9.13	&	1948.15	&$\pm$	805.41	\\
4959.0	&	[OIII]	&	202.60	&	160.19	&$\pm$	18.63	&	3790.43	&$\pm$	1526.28	\\
5006.9	&	[OIII]	&	714.05	&	547.79	&$\pm$	62.78	&	12497.30	&$\pm$	4969.05	\\
5875.6	&	HeI 	&	18.89	&	15.34	&$\pm$	1.44	&	207.30	&$\pm$	68.20	\\
6300.3	&	[OI]	&	7.61	&	6.12	&$\pm$	0.56	&	66.62	&$\pm$	20.36	\\
6312.1	&	[SIII]  	&	3.79	&	3.08	&$\pm$	0.28	&	33.36	&$\pm$	10.17	\\
6363.8	&	[OI]	&	...	&	...	&$\pm$	...	&	...	&	...	\\
6548.1	&	[NII]	&	11.31	&	8.99	&$\pm$	0.82	&	86.50	&$\pm$	25.38	\\
6562.9	&	HI	&	719.14	&	574.08	&$\pm$	52.61	&	5482.76	&$\pm$	1604.51	\\
6583.4	&	[NII]  	&	39.40	&	31.64	&$\pm$	2.90	&	299.19	&$\pm$	87.26	\\
6678.2	&	 HeI 	&	6.17	&	5.09	&$\pm$	0.47	&	45.92	&$\pm$	13.19	\\
6716.5	&	[SII]  	&	25.94	&	21.66	&$\pm$	2.01	&	191.60	&$\pm$	54.68	\\
6730.7	&	[SII]	&	19.80	&	16.60	&$\pm$	1.54	&	145.80	&$\pm$	41.51	\\
7065.3	&	HeI	&	...	&	...	&	...	&	...	&	...	\\
7135.8	&	[AIII]	&	16.76	&	15.81	&$\pm$	1.56	&	113.71	&$\pm$	30.31	\\
\hline
\end{tabular}
\end{minipage}
\end{center}
\end{table*}

Data were reduced following standard procedures. After correcting for atmospheric extinction and telluric
bands, we calibrated the observed fluxes using spectrophotometric
standards taken during the same night.  Keck fluxes were normalized to
VLT observations.  For the VLT spectra, we evaluate a relative flux
uncertainty of $\sim 2\%$ in the wavelength range 4100--7150\, \AA \
from the calibration of the spectrophotometric standards. The
uncertainty increases considerably outside this range of wavelengths,
reaching $\sim 16\%$ shortward of 4000\, \AA . For this reason, we do
not consider this region of VLT spectra in the following analysis. For
the Keck spectra a relative flux uncertainty of $\sim 5\%$ is
expected.

The entire analysis of the reduced spectra was carried out without any standard
package, in order to have more flexibility and control in the adopted
procedures. For the first night of observation we analyse each segment and then we
focus on the co-added spectrum. For the following nights, we compare the co-added spectrum to the one acquired on December 20: 
in this way we are able to detect any variability and spurious faint lines. We find that all of the identified 
lines are narrower than the
instrument resolution; hence we decided to fit a Gaussian of constant width to each emission line profile.
Using the four strongest emission features of each spectrum we derive a redshift $z=0.10536 \pm 0.00007$.
We believe our uncertainty to be dominated by systematics, since
line profiles are not strictly Gaussian. \cite{proc04} found 
$z=0.1055 \pm 0.0001$.  The host galaxy of GRB\,031203 therefore belongs to the local universe: with a luminosty
distance $D_{\mathrm{L}}=473\,\mathrm{ Mpc}$, GRB\,031203 is one of closest known long-duration bursts, along with the anomalous GRB\,980425 
\citep{Galama}, GRB\,030329 and GRB\,060218. 

Our measures are summarized in Tables ~\ref{Tab:Obs1} and
\ref{Tab:Obs2}. We list observed fluxes, extinction-corrected fluxes,
and equivalent widths of emission lines with certain detection and
identification. Observed-flux uncertainties are dominated by errors
in the flux-calibration procedure and in the subtraction of the
underlying continuum.  For extinction-corrected flux uncertainties,
we also consider errors coming from the extinction-correction
procedure (see Sect. \ref{Sec:Ext}).

\section{Galactic and intrinsic extinction}
\label{Sec:Ext}
The aim of this section is to estimate the extinction correction
necessary for calculating the total line fluxes of the observed
emission lines. First, we try to estimate the Milky Way absorption;
second, we derive the host-galaxy extinction correction using Balmer
decrements.

The Galactic extinction correction is not easy to assess because of
the location of GRB\,031203: the burst is very close to the
Galactic plane, at $l=255^\circ$, $b=-4.6^\circ$ (see \citealt{Cobb}
for detailed astrometry of HG\,031203). According to the far-IR map by
\cite{Schlegel}, the sight line to HG\,031203 has an inferred
reddening value $E_{\mathrm{MW}}(B-V)= 1.04$. However,
\cite{Dutra} argue that the far-IR analysis overestimates the real
$E_{\mathrm{MW}}(B-V)$ value by a factor of $\sim 25$\%.
Consequently, they suggest scaling the previous value by that factor.
Applying this correction, we get $E_{\mathrm{MW}}(B-V)= 0.78
$. However, the possibility of systematic
errors in the \cite{Schlegel} analysis, errors which could lead to an
underestimate of the reddening, has been considered by \cite{Willick}.
Moreover, \cite{Schlegel} invite the reader not to trust their
predicted reddenings of sight lines at low Galactic latitudes
($|b|<5^\circ$) because of the unresolved temperature structure of our
Galaxy and the presence of unremoved contaminating sources in their
map. The sight line to HG\,031203 falls in this category. Because of
this, we need other independent estimates of the Galactic absorption.

First, \cite{Dutra} find a spectroscopic reddening
$E_{\mathrm{MW}}(B-V)= 0.65$ for ESO\,314-2.  With a
Galactic longitude $l=261.5^\circ$ and latitude $b=4.1^\circ$, this
galaxy is nearly symmetric to HG\,031203 with respect to the Galactic
plane. Since we do not expect the interstellar medium (ISM) to be
highly clumped and irregularly distributed at this longitude, it seems
reasonable to assume that the two objects are affected by a similar
Galactic extinction. The mean reddening value found so far is then
$E_{\mathrm{MW}}(B-V) \approx 0.72$.  Second, thanks to
the NASA Extragalactic Database (NED), we gathered information about the Galactic absorption which affects
objects with lines of sight near that of HG\,031203. We found four
galaxies in a radius of $10$\arcmin\ from the host, with a mean
$E_{\mathrm{MW}}(B-V) \approx 0.72$. Consequently, we
consider $E_{\mathrm{MW}}(B-V) \approx 0.72$ to be a good
estimate of the Galactic absorption in the direction of the burst.

Adopting $E_{\mathrm{MW}}(B-V) \approx 0.72$, we
estimate for each spectrum the intrinsic reddening through comparison
of Balmer-line decrements under the assumption of Case B recombination
\citep{OsFerl}. We assumed a \cite{Cardelli} extinction law with
$R_{\mathrm{V}} = 3.1$ for both our Galaxy and HG\,031203.  The use of
a Small Magellanic Cloud extinction law for the host would not
significantly affect the results we obtain.  Table ~\ref{first} shows,
for each night of observation, the observed Balmer line ratios and the
intrinsic reddening $(E_{\mathrm{HG}}(B-V))$ derived from
them. The errors are calculated by simply propagating the
uncertainties of the measured Balmer ratios. Also shown is the
reddening value estimated from \cite{proc04} data. We re-analysed 
their spectrum in order to have a homogeneous dataset spanning the
longest possible temporal range.
In particular, the
intrinsic extinction correction we derive for this spectrum is higher
than the value \cite{proc04} found ($E_{\mathrm{HG}}(B-V)=0.39$).
We believe that this is mainly related to the different Galactic absorption
assumed by them in the direction of the burst. A minor role is  
played by their use of a single extinction law, a procedure that neglects 
the limited effects
caused by the non-zero redshift of the host galaxy.

We are now able to apply the derived extinction corrections to the
observed emission-line fluxes listed in Table \ref{Tab:Obs1} and Table
\ref{Tab:Obs2}.  In the same tables, the
extinction-corrected
fluxes and errors are also reported. 
From the badly constrained
Galactic extinction we expect an additional uncertainty of $\sim15\%$,
ranging from 10 to $25\%$ at 9000 and $4000\,\AA$ respectively.
Extinction and $K$-corrected absolute
magnitudes of HG\,031203 are listed in Table \ref{Tab:Mag}.

\begin{table}
\begin{minipage}[t]{\columnwidth}
\caption[Table]{HG\,031203 intrinsic reddening values and hydrogen Balmer line ratios from which it has been derived. 
Reliable observed emission line fluxes has been used only. We assumed $E_{\mathrm{MW}}(B-V)= 0.72$.  }
\label{first}
\begin{center}
\begin{tabular}{lllllll}
\hline
Date &  $\mathrm{H\alpha /H\beta}$ & $\mathrm{H\beta/H\gamma}$ & $\mathrm{H\gamma /H\delta}$ & \multicolumn{2}{c}{$E_{\mathrm{HG}}(B-V)$ }\\
\hline
\hline
03/12/06 \footnote{Re-analysis of the \cite{proc04} spectrum.} & 8.942 & 3.727 & 2.623 & 0.5516  &$\pm$0.0167  \\
03/12/20  & 7.937 & 3.662 & 2.391 & 0.4329  &$\pm$0.0471 \\
03/12/30  & 7.449 & 3.712 & ----- & 0.3843  &$\pm$0.0446  \\
04/03/02  & 7.451 & 3.803 & 2.217 & 0.3806  &$\pm$0.0874 \\
04/03/16  & 8.330 & 3.166 & -----  & 0.3846  &$\pm$0.1904 \\
04/09/20  & 8.086 & 3.697 & 2.293 & 0.4548  &$\pm$0.0467 \\
\hline
\end{tabular}
\end{center}
\end{minipage}
\end{table}
\section{Nebular parameters and metallicity}
\label{Sec:Met}

\begin{table}
\begin{minipage}[t]{\columnwidth}
\caption[Table]{Diagnostic line intensity ratios for HG\,031203.  }
\label{Tab:DiagRatio}
\begin{center}
\begin{tabular}{l|ll}
\hline
 &  $([\mathrm{OIII}]_{5007})/(\mathrm{H\beta}_{4861})$ & $([\mathrm{NII}]_{6583})/(\mathrm{H\alpha}_{6563})$\\
\hline
\hline
VLT\footnote{VLT averaged value.} & 0.80&-1.28 \\
\hline
&  $([\mathrm{SII}]_{6716+6731})/(\mathrm{H\alpha}_{6563})$ & $([\mathrm{OI}]_{6300})/(\mathrm{H\alpha}_{6563})$\\
\hline
\hline
VLT & -1.50& -2.00\\
\hline
\end{tabular}
\end{center}
\end{minipage}
\end{table}

In this section we first derive the physical features of the nebular
region  within HG\,031203: emission-lines ratios, electron temperature,
and electron density. Then we examine the chemical abundances of the
elements.

Table ~\ref{Tab:DiagRatio} lists the diagnostic line-intensity ratios
corrected for reddening obtained from VLT spectra taken on 2003
Dec. 20th, 2003 Dec. 30th, 2004 March 2nd, and 2004 Sep. 20th. The
line-intensity ratios for emission-line galaxies have been extensively
discussed in the literature; in particular, adopting the samples of
HII regions and active galactic nuclei (AGNs) presented by
\cite{OsFerl} as reference samples, it is easy to see that all of the
ratios obtained for HG\,031203 are typical of starburst galaxies, in
agreement with \cite{proc04}. This fact, along with the
observation that all of the lines are narrower than the instrumental
resolution, allows us to exclude the presence of an AGN as source of
ionization. Moreover, it establishes a strong link between HG\,031203
and star formation; the emission lines are produced by HII regions .

Following the standard nebular theory discussed by \cite{OsFerl}, we
know that the electron density may be assessed by observing the
effects of collisional de-excitation; while the electron temperature
may be determined from measurements of intensity ratios of lines
emitted by a single ion from two levels having considerably different
excitation energies. From the $\mathrm{[S~II]}$ and $\mathrm{[O~III]}$
emission-line ratios averaged over the four VLT spectra, we find for
HG\,031203 an electron density $N_{\mathrm{e}} = 160 \pm
20\,\mathrm{cm^{-3}}$ and a temperature $T_{\mathrm{e}}=12400 \pm
100\,\mathrm{K}$ (quoted error bars include only the measurement
uncertainties). An additional uncertainty of the order of $2\%$
is expected to affect our electron temperature estimate
because of the badly constrained Galactic color excess.
However, we
note the substantial agreement between our results and those of
\cite{proc04} assuming a two-zone model for the temperature: for the
moderate zone they derive $T_{\mathrm{mod}}=13\,400 \pm 2000\,
\mathrm{K}$, and assuming $T_{\mathrm{low}}=12\,900\,\mathrm{K}$ they
find $N_{\mathrm{e}}\approx 300\,\mathrm{cm^{-3}}$ (the value adopted
for both zones).

Next, we estimate the relative elemental abundances for He, N, O, Ne, S, Ar following the procedure detailed
in \cite{OsFerl}. Atomic parameters (recombination and collision excitation coefficients) have been
interpolated to the ISM temperature and density estimated above. Ionization correction factors of \cite{Izotov} have been used
in order to account for unobserved stages of ionization, while the solar reference values come from \cite{OsFerl}.
Our results are shown in Table ~\ref{table2}. 

\begin{figure}[h]
\begin{center}
\includegraphics[bb= 0 0  310 230,scale=0.78]{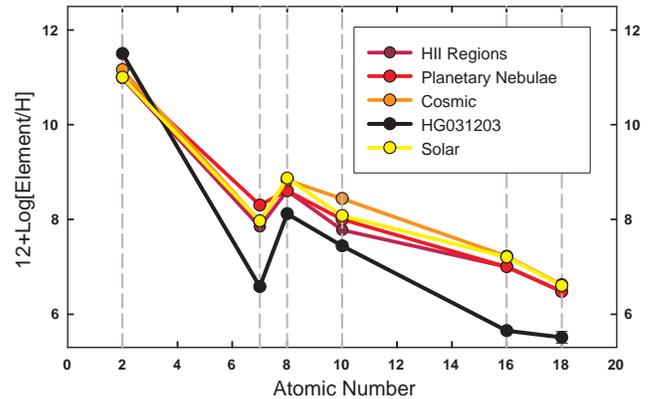}
\caption{Relative elemental abundances as function of the atomic number. Reference values for other environments taken from
\cite{OsFerl}.}
\label{Figure41}
\end{center}
\end{figure}

For each element we report the weighted-average value over the four
VLT spectra; uncertainties were calculated using the standard theory
of error propagation. The badly constrained Galactic color excess
is source of systematic errors estimated to be $0.03$ dex at most.
Figure ~\ref{Figure41} shows the relative elemental abundances of
HG\,031203 as a function of the atomic number, $Z$. Solar and cosmic
abundances, and abundances of H~II regions and planetary nebulae are
also plotted for comparison. From this figure, it is clear that
GRB\,031203 occurred in a low-metallicity environment. 
We will quantify this statement in Sect. ~\ref{Sec:Disc}.
Here we only note that HG\,031203 is characterised by
$12+\rm{log}(O/H)=8.12\pm0.04$.
To this uncertainty, a further error of 0.03 dex needs
to be added, to account for the uncertainty in the dust correction.

\begin{table*}[!th]
\begin{minipage}[t]{\columnwidth}
\begin{center}
\caption[smallcaption]{Relative elemental abundances for HG\,031203.}
\label{table2}
\centering
\begin{tabular}{l|cccccc}
\hline
Element & He & N & O & Ne & S & Ar \\
\hline
Abundances & 11.50$\pm $0.04 & 6.58$\pm $0.06 & 8.12$\pm $0.04 & 7.44$\pm $0.06 & 5.65$\pm $0.04 & 5.51$\pm $0.12 \\
\hline
\end{tabular}
\end{center}
\end{minipage}
\end{table*}

\section{Star formation}
\label{Sec:SFR}

With a redshift $z = 0.1054$, HG\,031203 has a luminosity distance
$D_{\mathrm{L}}=473\,\mathrm{Mpc}$. According to \cite{Kennicutt},
the $\mathrm{H\alpha}$ luminosity derived from Tables \ref{Tab:Obs1}
and \ref{Tab:Obs2} and corrected for slit losses corresponds to a star-formation rate
$\mathrm{SFR(H\alpha)} = 12.9 \pm 2.2\,\mathrm{ M_{\sun}\,yr^{-1}}$
(value averaged over the four VLT spectra). 
The uncertainty arising
from the badly constrained Galactic color excess is estimated 
to be $\sim13\%$.

While we caution the reader that the uncertainty reported here does
not account for the uncertainties in the SFR calibration, estimated to
be $\sim30\%$ \citep{Kennicutt}, we also stress the
substantial agreement between the \cite{proc04} results and ours. For
HG\,031203 they find
$\mathrm{SFR(H\alpha)}=11\,\mathrm{M_{\sun}\,yr^{-1}}$, a
lower limit to the total SFR because of slit losses and the possible
presence of regions of star formation enshrouded in dust. 
Parenthetically, we note that
\cite{Watson} derived from X-observations an upper limit to the
HG\,031203 SFR of $150\pm 110\,\mathrm{M_{\sun}\,yr^{-1}}$.
   
We can further characterize the star-formation activity of HG\,031203
through the analysis of two other parameters: the rest-frame
equivalent width of the $\mathrm{H\alpha}$ emission line and the
specific star-formation rate (SSFR).  The galaxy is notable for its
rather strong $\mathrm{H\alpha}$ emission; we measure
$EW_{\mathrm{rest}}(\mathrm{H\alpha})=760\pm80\,\AA$ after the underlying
supernova continuum has faded away. We will contrast the properties of
HG\,031203 with local samples of galaxies in Sect. \ref{Sec:Disc};
however, here we note that this is an extremely high value, significantly
larger than that of normal star-forming galaxies in the local universe.

Another valuable diagnostic of star formation is the SSFR, here
defined as the ratio of $\mathrm{H\alpha}$ to $B$-band luminosity. For
HG\,031203 we measure $M_{B}=-21.00\,\pm\,0.39\, \mathrm{mag}$ (see
Table \ref{Tab:Mag}).  With an extinction-corrected $\mathrm{H\alpha}$
luminosity of $1.6 \times 10^{42}\,\mathrm{erg\, s^{-1}}$, HG\,031203 is
then characterised by $\mathrm{SSFR} \approx 0.1$, much higher than the
average SFR per unit $B$-band luminosity of ordinary local galaxies.

We will quantitatively discuss this results in Sect. \ref{Sec:Disc}.
\section{A WR bump?}
\label{Sec:WR}

\begin{figure}[!h]
\centering
\includegraphics[scale=0.6]{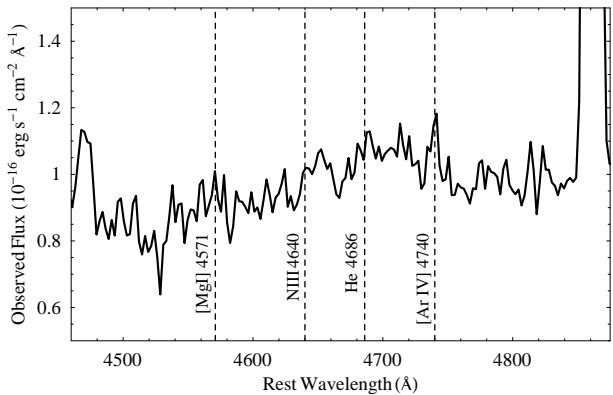}
  \caption{Result of the stacking of the host galaxy spectra taken on March 2nd and September 20th,
shown close to the region supposed to be populated by WR lines.}
  \label{Fig:WRSommato}
\end{figure}

In this section we investigate the spectra of HG031203 for possible signatures
of Wolf-Rayet (WR) stars emission, as previously claimed by \cite{Hammer}.
We pay particular attention to the presence of broad emission at 4686\,\AA \ due to HeII,
the strongest line in the optical for WR stars along with 4650\,\AA \ CIII and 5808\,\AA \ CIV
\citep{ContiMassey}. Furthermore, we investigate the spectra for the 
presence of other WR features, the most common of which is 4640\,\AA \ NIII. This is the 
only WR emission line, aside from 4686\,\AA \ HeII which is clearly identified in several
WR galaxies \citep{Conti}. Our results are shown in Fig. \ref{Fig:WRSommato}: in order to 
increase the signal to noise ratio and better address the potential presence
of WR signatures, we stacked our latest spectra (the spectra taken on December 20th and 30th
were significantly contamined by the SN emission).
From this figure it is clear that the identification of the HeII and 4640\,\AA
\ NIII emission lines is uncertain.

\section{Continuum emission}
\label{Sec:Cont}  
\begin{figure*}[!t]
\begin{center}
\includegraphics[scale=0.85]{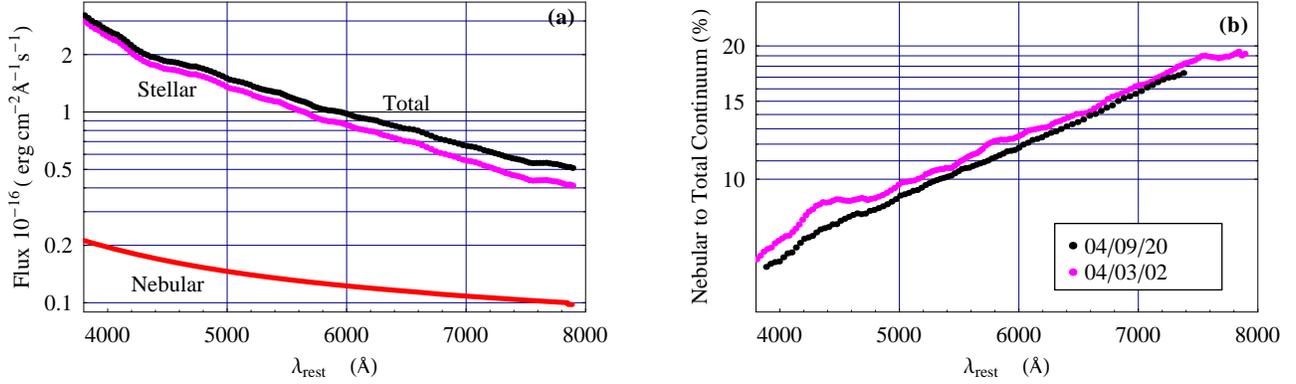}
\caption{(a) Nebular continuum, stellar continuum and total continuum of HG\,031203 obtained from
VLT observations taken in March 2004. We find analogous results for the host continuum observed in September
2004. (b) Ratio of nebular to total continuum for the VLT observations acquired in March and 
September 2004. Extinction corrected data are shown.}
\label{Fig:Cont}
\end{center}
\end{figure*}

So far we focussed on the line radiation emitted by HG\,031203. In
this section we analyse the continuum-emission properties of the host,
with the aim of further characterizing the large-scale environment of
the nearby GRB event.

The observed continuum consists of three components: stellar,
supernova, and nebular. We have no evidence for a nonthermal source;
the measured diagnostic line-intensity ratios corrected for reddening
(see Table \ref{Tab:DiagRatio}) clearly indicate that HG\,031203 is an
object dominated by photoionization, with line ratios typical of
starburst or H~II galaxies.

In order to exclude the SN contribution, we restrict our analysis to
the VLT spectra taken in March and September 2004. For the nebular
continuum, we identify two different sources of emission: the
recombination process and the two-photon decay of the $2^{2}S$ level
of hydrogen. Recombination processes lead to the emission of a rather
weak continuum in free-bound and free-free transitions. Since hydrogen
is the most abundant element, the H~I continuum is the strongest; on
the other hand, since the He~II lines are missing in HG\,031203
spectra, we expect negligible contribution from this ionization
stage. Recombination emission coefficients and emission coefficients
of two-photon decay have been interpolated to the interstellar medium
(ISM) parameters calculated above \citep{OsFerl}. We then normalized
the theoretical nebular continuum to the observed He~I and H~I line
fluxes (Tables \ref{Tab:Obs1} and \ref{Tab:Obs2}). Results are shown
in Fig.  \ref{Fig:Cont}. Since the nebular spectrum is redder than the
stellar spectrum, a bluer spectrum results if the nebular contribution
is removed.

In order to obtain more information about the HG\,031203 stellar
populations, we analyse the continuum contribution that is stellar in
origin. We adopt a simplified procedure: a detailed model, beyond the
scope of this paper, is in progress. Moreover, we handle a
limited portion of the host spectral energy distribution (SED). An
ultraviolet spectrum would be of particular interest to investigate
the ionizing continuum of the galaxy. We adopt a \cite{Kroupa} initial
mass function (IMF).
Next, we assume for simplicity that the observed stellar continuum is
due to main-sequence stars characterized by black-body emission, the 
main-sequence mass-temperature relation, and the main-sequence mass-radius
relation. As a consequence, according to our simplified model, the
host spectrum can be written as:

\begin{equation}
\label{Eq:ContMod}
F_{\rm{\lambda obs}}=\frac{(1+z)^{-1}}{4\pi D_{\rm{L}}^{2}}\int_{M_{\rm{i}}}^{M_{\rm{f}}}\xi (m)BB_{\lambda}(T(m))4\pi r^{2}(m)dm
\end{equation}
where:
\begin{equation}
\label{Eq:lambda}
\lambda=\frac{\lambda_{\rm{obs}}}{(1+z)}
\end{equation}
\[
      \begin{array}{lp{0.8\linewidth}}
       F_{\rm{\lambda obs}}& \textrm{Flux per unit observed wavelength} \\
	   D_{\rm{L}}& \textrm{Luminosity distance corresponding to a redshift $z$ }\\
	   \xi (m)& \textrm{IMF}\\
       BB_{\lambda}(T(m))& \textrm{Black body energy distribution at temperature $T$}\\ 
       r & \textrm{Stellar radius}\\
       M_{\rm{i}},M_{\rm{f}} & \textrm{Lower and upper end of the mass distribution. We use $M_{\rm{i}} =\, 0.08 \,\rm{M_{\sun}}$; $M_{\rm{f}}\, =$ free parameter of the model.}
      \end{array}
   \]
   
A $(1+z)^{-1}$ term is needed in order to account for the cosmological frequency or wavelength folding.
Finally, we fit the stellar continuum leaving the IMF normalization and 
the upper end of the mass distribution as free parameters of our model. A best-fit HG\,031203 SED
is derived. 

To improve the sensitivity of our model to the unobserved ionizing
continuum of the host, we consider the following complementary
approach. From the standard theory of recombination, well described by
\cite{OsFerl}, it is also possible to show that:

\begin{equation}
\label{Eq:FluxHBeta}
F(\rm{H\beta})=\frac{\rm{\alpha_{H\beta}}^{\rm{eff}}(\rm{H,T})}{\alpha_{B}(\rm{H,T})}\int_{0}^{\lambda_{0}}\frac{\lambda}{\rm{\lambda_{H\beta}}}{F_{\lambda}}d\lambda
\end{equation}
where:
\[
      \begin{array}{lp{0.8\linewidth}}
       F(\rm{H\beta}) & \textrm{Emission line flux of the hydrogen Balmer $\beta$ line}\\
       F_{\lambda }& \textrm{Flux per unit  wavelength estimated from $F_{\mathrm{\lambda obs} }$, Eq.(\ref{Eq:ContMod})} \\
	   \alpha_{\rm{H\beta}}^{\rm{eff}}(\rm{H,T}) & \textrm{Effective recombination coefficient for the hydrogen Balmer $\beta$ line at temperature $T$  \citep{OsFerl}}\\
	   \alpha_{\rm{B}}(\rm{H,T})& \textrm{Recombination coefficient for hydrogen at temperature $T$ \citep{OsFerl}}\\
	   \lambda_{0}& \textrm{Ionization threshold}. \\
      \end{array}
   \]
   
\begin{figure}[h]
\begin{center}
\includegraphics[scale=0.9]{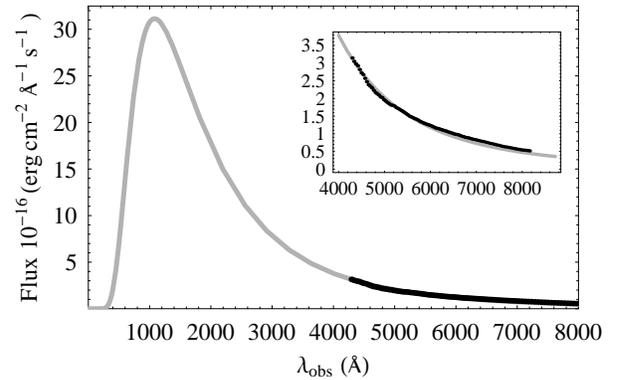}
\caption{Grey line: theoretical stellar spectral energy distribution of the host galaxy of GRB\,031203. 
The stellar continuum estimated from the observations taken in September 2004 is also shown 
for camparison (black line).}
\label{Fig:ContSpec}
\end{center}
\end{figure}

   Comparing the predicted $\mathrm{H\beta}$ flux (estimated using
the previous relation) to the observed $\mathrm{H\beta}$ flux (from
Table \ref{Tab:Obs1}), and then normalizing the free parameters of our
model to the observed $\mathrm{H\beta}$ flux, we finally derive:

\begin{equation}
 \left\{ \begin{array}{ll}
M_{\rm{f}}&\approx 25 \,\rm{M_{\sun}}\\
M_{\rm{tot}}&\approx\, 10^{9}\,\rm{ M_{\sun}}
\end{array}\right. 
\end{equation}

\noindent where $M_{\rm{f}}$ is the upper end of the stellar mass distribution, while $M_{\rm{tot}}$ is the 
total stellar mass of the emitting volume we observe. Our results are shown in Fig. \ref{Fig:ContSpec}.
Given the simplified method used, we expect our results to be reliable
within a factor of two. 
\section{Discussion}
\label{Sec:Disc}  
Our analysis of HG\,031203 reveals a small, metal-poor star-forming
galaxy. In this section we will discuss the physical properties of the
host, contrasting its characteristics with a complete sample of
$\sim$1,000 local $(z \leq 0.1)$, normal, star-forming galaxies from
KISS \citep{Salzer}.

In order to carry out a statistically significant analysis, the largest sample
of local LGRB HGs with associated SNe will be considered. This sample
includes HG\,980425, HG\,030329, HG\,031203, and HG\,060218.  We 
summarize the observed properties of the hosts in Table \ref{Tab:hosts}.

Recently, two additional LGRBs have been discovered at low redshift:
GRB\,060505 at $z = 0.09$ \citep{Fynbo} and GRB\,060614 at $z = 0.125$
\citep{DellaValle,Gal06}. However, these bursts were somewhat
peculiar, not showing any SN component down to deep limits, and there
is intense debate concerning their origin
(e.g., \citealt{Gehrels,King}).  We therefore chose to exclude their
host galaxies from the present analysis.

\begin{table}
      \caption[]{Properties of the LGRB/SN host galaxies of the local universe.}
         \label{Tab:hosts}
    $ 
         \begin{array}{lrrrrrrr}
            \hline
            \noalign{\smallskip}
             &\mathrm{HG\,980425} &\mathrm{ HG\,030329} &\mathrm{HG\,031203}& \mathrm{HG\,060218} \\
            \noalign{\smallskip}
            \hline
            \noalign{\smallskip}
            z &	0.0085^{\mathrm{f}}& 	0.1685^{\mathrm{f}}&	0.10536^{\mathrm{h}}&	0.0335^{\mathrm{d}}\\
            12+\log{(\mathrm{O/H})}& 8.39^{\mathrm{a}}  & 7.8^{\mathrm{c}}   & 8.12^{\mathrm{h}} &8.0^{\mathrm{d}} \\
            & 8.25^{\mathrm{b}}  & \\
			M_{B}\,\mathrm{mag} &-17.65^{\mathrm{f}}  &	-16.5^{\mathrm{g}} &	-21.00^{\mathrm{h}}  &	-15.86^{\mathrm{d}} 	\\
			\mathrm{SFR(H\alpha)}\,\mathrm{M_{\sun}\,yr^{-1}}&0.35^{\mathrm{f}}  &	0.40^{\mathrm{f}} & 12.32^{\mathrm{h}}   &	>0.07^{\mathrm{e}} 	\\
			\mathrm{SSFR} & 0.05&	0.19&	0.1&	>0.04\\
            \noalign{\smallskip}
            \hline
         \end{array}
     $ 
\begin{list}{}{}
\item[$^{\mathrm{a}}$] WR Region \cite{Hammer}, $T_{e}$ method.
\item[$^{\mathrm{b}}$] SN Region \cite{Hammer}, $T_{e}$ method.
\item[$^{\mathrm{c}}$] \cite{Thoene}, $R_{23}$ method.
\item[$^{\mathrm{d}}$] \cite{Modjaz}, $R_{23}$ method.
\item[$^{\mathrm{e}}$] \cite{Ferrero}.
\item[$^{\mathrm{f}}$] \cite{Sollerman05}.
\item[$^{\mathrm{g}}$] \cite{Gorosabel}.
\item[$^{\mathrm{h}}$] This work.
\end{list}
   \end{table}
\subsection{Intrinsic extinction}
\label{SubSec:IE}

\begin{figure}[h]
\begin{center}
\includegraphics[scale=0.7]{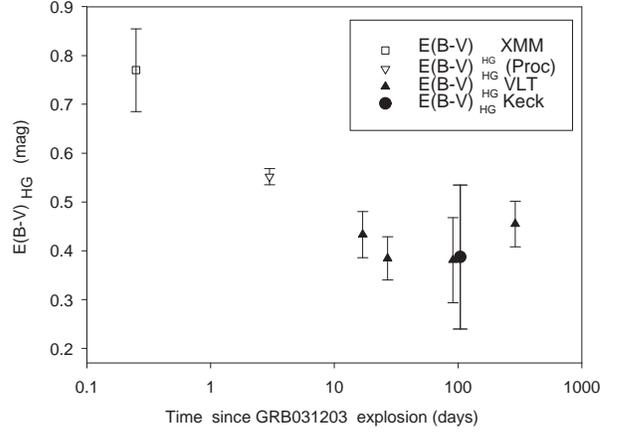}
\caption{Intrinsic reddening value $E_{\mathrm{HG}}(B-V)$ for HG\,031203 as a function of time elapsed from GRB explosion. 
While the first point is derived from X-ray observations, all the other values are estimated through comparisons
of Balmer line ratios, assuming $E_{\mathrm{MW}}(B-V)= 0.72$. See Table \ref{first} for the numerical values of the data shown.}
\label{Figure71}
\end{center}
\end{figure}

Here we gather all of the available information about
the intrinsic extinction of HG\,031203. Our goal is to test the
possibility that the extinction changed with time.
Fig.~\ref{Figure71} shows the data. In particular, the first point shown is
derived from XMM-Newton observations: from the X-ray data
\cite{Vaughan} estimate a hydrogen column density $N_{\mathrm{H}}=(8.8
\pm 0.5)\times 10^{21} \,\mathrm{ cm^{-2}}$. Assuming the standard
hydrogen column density vs. reddening correlation,
\begin{equation}
\label{rel:2}
E(B-V)=\frac{N_{\rm{H}}}{5.9\times 10^{21}~{\rm cm}^{-2},}
\end{equation} 
\noindent
we derive a total reddening $E(B-V)_{\mathrm{TOT}}=1.49 \pm
0.08$ in the direction of the burst. Adopting a single
extinction law and $E(B-V)_{\mathrm{MW}}=0.72$, this
would formally correspond to an intrinsic reddening
$E(B-V)_{\mathrm{HG}}=0.77 \pm 0.08$.  The second point
in the plot is derived from the re-analysis of the \cite{proc04} spectrum.

The X-ray observations started only 6 hours after the burst, while the
last VLT spectrum is from $\sim 9$ months after the
explosion.  On the other hand, the value derived from the hydrogen
column density strongly depends on the assumed dust-grain size and 
gas-to-dust ratio. In particular, increasing one of these two parameters,
we would obtain a smaller $E(B-V)/N_{\mathrm{H}}$ ratio and a larger
$R_{V}$ value \citep{Maiolino}.  The gas-to-dust ratio is expected to
have substantial variations in the outskirts of the Galactic plane
\citep{Burton}, regions that the X-ray signal from GRB\,031203 had to
cross. For this reason we are skeptical about the reliability of the
intrinsic reddening estimate derived from X-ray observations.

Apart from the X-ray data point, there seems to be a tendency of the
HG\,031203 reddening value to decline with time.  However, we must
also consider that this conclusion relies completely on the
reliability of the value of $E(B-V)_{\mathrm{HG}}$ derived from
our analysis of the \citealt{proc04} spectrum. Our VLT
$E(B-V)_{\mathrm{HG}}$ estimates are consistent with no time evolution
of the parameter. Moreover, the
Balmer decrements we used to calculate the extinction corrections are
particularly sensitive to small errors in the line-fitting
procedure. These factors seem to cast doubt on the variable
intrinsic reddening hypothesis. 

\begin{figure}[h]
\begin{center}
\includegraphics[scale=0.80]{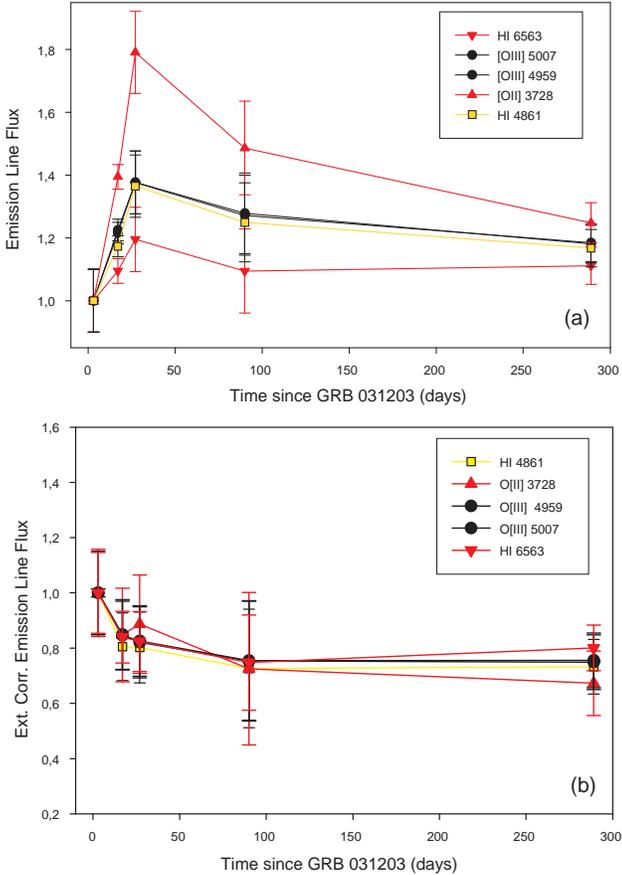}
\caption{Evolution of the five brigthest emission lines fluxes.
Emission line fluxes have been normalized to the observations taken on December 6th (re-analysed
\citealt{proc04} spectrum).Data have been corrected for 
the different slit losses.
(a): Non extinction corrected data;
(b): Extinction corrected data.}
\label{Fig:LineEvolution}
\end{center}
\end{figure}

In order to better address this issue we considered the potential
evolution with time of emission line fluxes. The observed line
fluxes are expected to evolve because of the joint action
of the gradual ionization of the medium by the GRB radiation 
(we refer the reader to \citealt{Perna1998}; \citealt{Perna2000} 
and references therein) and
a time-dependent intrinsic extinction. 
Given the present set of data we can check this hypothesis on a 
month time scale. 
To this end,
it is necessary to correct the emission line fluxes
for slit losses. A relative correction of $\approx 10\%$
has been applied to our VLT fluxes to account for this fact.
Before applying the extinction corrections (Fig. \ref{Fig:LineEvolution}a), we find that 38\% of the emission lines with
firm identification shows no time evolution
at $1\sigma$ confidence level; 71\% at $2\sigma$ and 81\% at $3\sigma$. 
Correcting
for the Galactic and the intrinsic reddening (using the different $E_{HG}(B-V)$ 
values listed in Table \ref{first}) the situation changes as follows (Fig. \ref{Fig:LineEvolution}b):
57\% of the lines are consistent with the no-evolution hypothesis at $1\sigma$;
95\% at $2\sigma$ while all the emission lines show no-evolution at
$3\sigma$. Non-corrected fluxes show variations of a factor of 2 at most,
(see Fig. \ref{Fig:LineEvolution})
with the most consistent evolutions detected at the shortest wavelength and 
with the peak of the variation occurring approximately 30 days after
the GRB explosion (epoch at which \citealt{Malesani} detect the peak of the SN2003lw light curve).
While the wavelenght-dependence implicitly suggests the time evolution
of the intrinsic color excess to be the source of the limited flux
evolution we detect, we conclude the absence of compelling observational evidence
of emission line fluxes evolution with time on a typical month-scale.

Hence, the detection of the ``cleaning 
action'' of the burst remains at low significance level.
 
\subsection{Metallicity}
\label{SubSec:Met}
With $12+\log{\mathrm{(O/H)}}=8.12$, HG\,031203 has a low metallicity
(Fig. ~\ref{Figure41}).  In order to quantify this result, we contrast
the hosts properties against a sample of ordinary star-forming
galaxies. For the KISS galaxy sample, \cite{Melbourne} find a
metallicity vs. $B$-band luminosity relation expressed by
\begin{equation}
\label{Eq:MetLum}
12+\log{\mathrm{(O/H)}}=(4.059\pm 0.17)-(0.240\pm 0.006)M_{B},
\end{equation}
\noindent
with a root-mean-square (rms) deviation of residuals of
$0.252\,\mathrm{dex}$. 
Metallicities have been estimated
by  \cite{Melbourne}
using the $T_{\rm{e}}$ method. Where not possible, the authors used the standard
$R_{23}$ method improved by the introduction of the $p_{3}$
factor (\citealt{Pilyugin}). In this way the $R_{23}$
method was found to correlate with the $T_{\rm{e}}$
method to within 0.1 dex for starburts with metallicities
below $12+\rm{log}(O/H)=7.9$. Results obtained in this way 
were used to calibrate relations between metallicity and emission-lines ratios.

According to the above relation, with $M_{B}
\approx -21$ (Sect. ~\ref{Sec:SFR}) HG\,031203 would have
had a relative oxygen abundance $12+\log{\mathrm{(O/H)}}\approx 9$,
$0.9\,\,\mathrm{dex}$ above the value we measure. Actually, the host
metallicity is lower than that of more than 99\% of all KISS galaxies with
$M_{B}\leq -21$.

\begin{figure}[h]
\begin{center}
\includegraphics[bb= 0 80 320 307,scale=0.75]{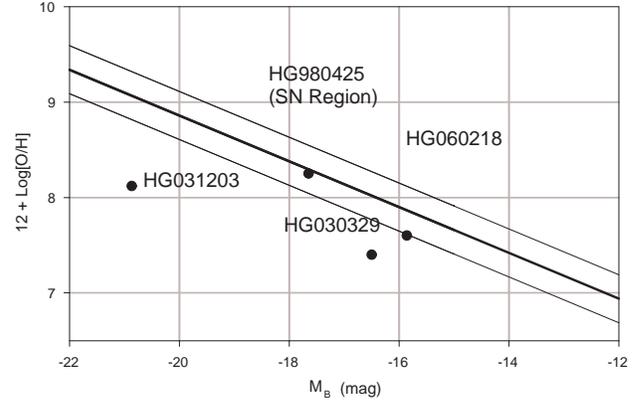}
\caption{Metallicity-luminosity relationship for KISS galaxies (thick line) and for local LGRB HGs.
For HG\,030329 and HG\,060218 we applied the metallicity  conversion
of L.J. Kewley \& S.L. Ellison (2007, in preparation)  to the $R_{23}$ estimates of Table \ref{Tab:hosts}.
Thin lines mark the rms 0.25 dex deviation of residuals about the fit. We refer the reader to
\cite{Melbourne} for details. }
\label{Fig:KISSMet}
\end{center}
\end{figure}

Fig. ~\ref{Fig:KISSMet} offers a graphic view of this offset,
interpreted by \cite{proc04} as a signature of a very young 
star-forming region. These authors speculate that HG\,031203 has been
observed prior to the production and/or distribution of metals into
its nebular regions. From the same figure it is also clear that this
is not a property shared by all the HGs of the local universe with 
associated SN: HG\,060218 and HG\,980425 follow the KISS trend
within the scatter, while HG\,030329 falls outside if we prefer
the lower branch metallicity estimate as indicated by \cite{Stanek}
and \cite{Thoene}.
We therefore do not find a compelling evidence for all the HGs to be
characterised by a substantially lower metallicity.

A similar conclusion has been independently derived by\cite{Wolf} analysing a subsample of the
~\cite{Fruchter06} data and considering only HGs in the redshift interval
$0.2-1$. These authors find all of the hosts to be within $\pm
1/3\,\mathrm{dex}$ of the metallicity-luminosity relation by
\cite{Kobulnicky} and conclude that there is no metallicity bias
within the statistical significance of five objects. Using a large sample
of local luminous star-forming
SDSS galaxies previously studied by \cite{Tremonti} as a reference
sample, \cite{Stanek} conclude that local LGRB/SN hosts have
oxygen abundances much lower that would be expected if local GRBs
traced local star formation independently of metallicity.
Indeed, a larger
and more representative sample of LGRB hosts with well-assessed
physical properties is needed before being able to definitively
answer the host peculiarity question.


Metallicity seems to play a central role in the production of a LGRB
event: theoretical models favour progenitors of low
metallicity. 
Finding a dependence of the LGRB rate on host
metallicity would therefore be of particular interest.  \cite{Stanek},
after investigating the physical properties of LGRB/SN hosts with
$z\leq 0.2$, find a host-metallicity threshold for producing
cosmological GRBs of $12+\log{(O/H)}\approx 8.0$.

However, after developing a Monte Carlo code to generate LGRB events
within cosmological hydrodynamical simulations consistent with the
$\Lambda$CDM model, ~\cite{Nuza} find no tight correlation between
the O/H abundance of the LGRB progenitor star and the mean metallicity
of the HG. From an observational point of view, even spectroscopic but
spatially unresolved measurements of a mean host metallicity do not
directly reflect the metallicity of the immediate burst environment
\citep{Wolf}.

Moreover, uncertainties exist with respect to the calibration of the
various methods used to measure the oxygen abundances: direct
measurement from observed oxygen emission lines \citep{OsFerl};
$R_{23}$ indicator \citep{Pagel}; electron temperature
$(T_{\mathrm{e}})$ method (see, e.g., \citealt{Kennicutt03}); and the
recently developed $\mathrm{O~II_{\mathrm{RL}}}$ method \citep{Peimbert}.

Finally, it is not well established whether the oxygen abundance is
representative of other elements: different elements might provide
much of the opacity in stellar winds and atmospheres, affecting in
this way the stellar evolution and the possible LGRB production.

\subsection{Star Formation Activity of local LGRB HGs}
\label{SubSec:SF}

HG\,031203 is notable for its very intense star-formation
activity. First, this galaxy has $\mathrm{SFR}>10\,\mathrm{ M_{\sun}\,
yr^{-1}}$, greater than 98\% of all low-redshift ($z \leq 0.1$)
galaxies \citep{proc04,Nakamura}.  Second, HG\,031203 exhibits a
significant rest-frame equivalent width of the $\mathrm{H\alpha}$
line, increasing from 2003 December to 2004 September because of the
decreasing underlying SN\,2003lw continuum. Considering the SN
contribution in September to be negligible, we measure for HG\,031203
$EW_{\mathrm{rest}}(\mathrm{H\alpha})=760\pm80\,\AA$.

\begin{figure}[h]
\begin{center}
\includegraphics[scale=0.55]{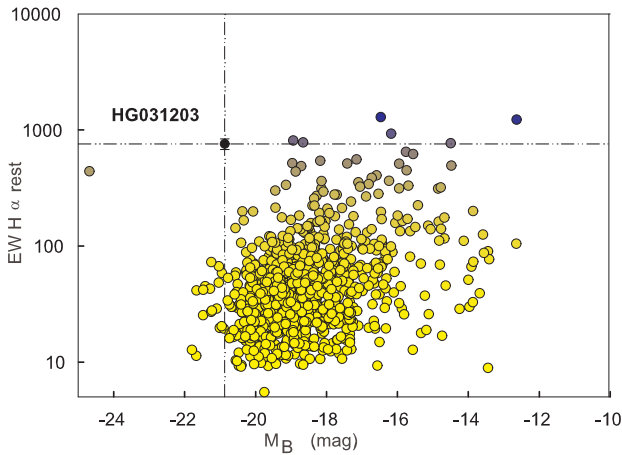}
\caption{Equivalent width at rest of the $\mathrm{H\alpha}$ line for HG\,031203 and for a complete sample 
of normal star forming galaxies of the local universe (KPNO International Spectroscopic Survey of galaxies, \citealt{Salzer}) as a function of 
absolute $B$-band magnitude. KISS data are not extinction corrected.}
\label{Fig:EW}
\end{center}
\end{figure}
Fig. \ref{Fig:EW} compares this value directly with the same quantity
measured for local, normal, star-forming galaxies from KISS.  It is
clear that HG\,031203 shows an $EW_{\mathrm{rest}}(\mathrm{H\alpha})$
value significantly larger than that of ordinary galaxies at the same
redshift; it is larger than that of 99\% of all galaxies in the
reference sample.

\begin{figure}[h]
\begin{center}
\includegraphics[bb=0 0 380 270,scale=0.65]{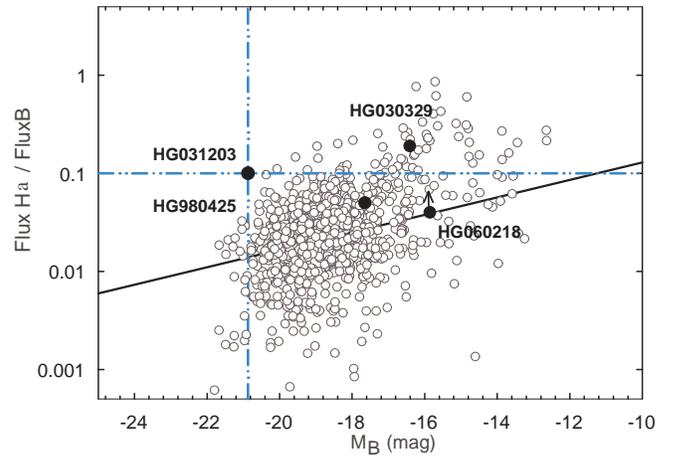}
\caption{SSFRs of local GRB hosts against a complete sample of normal star forming galaxies 
of the local universe (KISS galaxies) as a function of absolute $B$-band magnitude. KISS data are
 not extinction corrected and provide therefore upper limits to the parameter. Thick line: KISS data best fit line.}
\label{Fig:SSFR}
\end{center}
\end{figure}

Third, the specific star formation rate (SSFR) of the host is also remarkable. As before, we compare
the extinction corrected H$\alpha/B$-band luminosity of HG\,031203 against the KISS galaxies (Fig. \ref{Fig:SSFR}).
KISS data are not extinction-corrected; hence, they provide upper
limits to the H$\alpha$/$B$-band flux ratio.  In spite of this, HG\,031203 falls in the
upper end of the distribution and exhibits a SSFR greater than that
of 98\% of all KISS galaxies. Furthermore, if we consider the SSFR of
the largest sample of local LGRB/SN HGs (reference data are reported
in Table \ref{Tab:hosts}), we find that all of the hosts are in the
upper part of the panel.  This fact reveals a trend for LGRB/SN hosts
to be characterized by SSFRs which are larger than those of ordinary
star-forming galaxies at similar redshifts, as previously reported by
\cite{Christensen}.
\subsection{Wolf-Rayet emission in HG\,031203 spectra? }
\label{SubSec:WR}

A deep understanding of the GRB production mechanism requires us to
study GRB environments in detail.  Wolf-Rayet (WR) stars are the
favoured GRB progenitors in the collapsar model
\citep{Woosley99,Meszaros}; detailed stellar evolution models of WR
stars are consistent with the conditions for GRB production via
collapsar, as shown by \cite{Hirschi}. In particular, the evolution of
these stars is able to reproduce the conditions for black hole
formation and for the loss of the hydrogen-rich envelope (as is
expected if GRBs originate from SNe~Ib or Ic). Furthermore,
these stars are able to retain sufficient angular momentum to form an
accretion disk around the black hole. Finally, if only stars of the
particular subtype WO \citep{Conti} are considered, then the GRB
production rate can be reproduced \citep{Hirschi}.

Following \cite{Conti}, WR galaxies are defined to be those galaxies
whose integrated spectra show a broad ($\mathrm{FWHM} \geq10\, \AA$)
emission feature around 4650\,\AA, attributed to WR stars. Along with
the He~II $\lambda$4686 line, we know that C~III $\lambda$4650 C~IV
$\lambda$5808 are the strongest optical lines of WR stars while
N~III $\lambda$4640 is one of the most common feature of WR galaxies. 
Examples of typical WR-galaxy spectra are
shown by \cite{Guseva}.  Investigating the spectrum of the GRB\,031203
host taken in September 2004, \cite{Hammer} detect a blue bump around
the He~II $\lambda$4686 line, and conclude that WR stars are present
in the host galaxy. While we also detect an emission bump around the
expected position of the He~II line in each of the spectra we
analysed, we consider dubious the identification of the He~II
$\lambda$4686 and N~III $\lambda$4640 lines (see Fig. \ref{Fig:WRSommato}).
We do not exclude that this bump of emission might in principle arise
from blending of faint and broad emission lines emitted by WR stars;
however we emphasize that it is not possible to classify
the HG of GRB\,031203 as a
WR galaxy from the detected emission lines.

\section{Conclusions}
\label{Sec:Conc}
HG\,031203 offers a precious opportunity to study in detail the
environment of a nearby LGRB. Here we show that this galaxy is notable
for its star-formation activity, with
$EW_{\mathrm{rest}}(\mathrm{H\alpha})=760\pm80\,\AA$ and
$\mathrm{SFR(H\alpha)}=12.9\pm 2.2\,\mathrm{M_{\sun}\,yr^{-1}}$; we
believe these values to be signatures of the very intense burst of
star formation that is occurring in this galaxy. Furthermore, we
suggest that the peculiarity of LGRB HGs is that their star-formation
efficiencies (SSFR) are significantly higher than in ordinary
star-forming galaxies at the same redshift.

We do not find compelling
evidence for all the LGRB HGs of the local universe to have
a substantially lower metallicity when compared to 
normal star forming galaxies with the same $B$-band luminosity. However,
HG\,031203 is a clear outlier, with its
exceptionally low metallicity.

After searching for possible WR emission features in our spectra, we
conclude that the detection and identification of typical WR emission lines
is uncertain.

Finally, our data suggest, but do not clearly support, the
possibility that the intrinsic extinction of HG\,031203 evolved with
the time as a consequence of the exposure of dust grains to the hard
radiation of the GRB.

\begin{acknowledgements}
	We thank an anonymous referee for constructive criticism and useful suggestions.
      We are grateful to the ESO staff at Paranal for carefully performing all our observations
      and for many useful suggestions, as well as to the Keck
		staff for their assistance. We thank Jason Prochaska for giving us his reduced data.
      DM acknowledges the Instrument Center for Danish Astrophysics for support.
      The Dark Cosmology Centre is founded by the Danish Research National Foundation.
     	The data presented in this paper were obtained with ESO telescopes under programmes
	072.D-0480, 072.D-0137 and 073.D-0255 as well as
	with the W. M. Keck Observatory (a partnership between Caltech, the
	University of California, and NASA), which was made possible by the
	generous financial support of the W. M. Keck Foundation.	
	This research was
	financially supported by ASI grant I/R/039/04, MIVR grant 2005025417,
	and USA NSF grant AST--0607485, and NASA/Swift grant NNG06GI86G.
\end{acknowledgements}

\bibliographystyle{aa}
\bibliography{7698file}

\end{document}